\documentclass[aps,prl,twocolumn,showpacs,preprintnumbers,amsmath,amssymb]{revtex4-1}

\usepackage{graphicx}
\usepackage{dcolumn}
\usepackage{bm}
\usepackage[mathlines]{lineno}
\usepackage[urlcolor=blue, colorlinks=true]{hyperref}
\usepackage{textcomp}

\begin{document}
\title{Electron interference in ballistic graphene nanoconstrictions}

\author{Jens Baringhaus$^{1}$, Mikkel Settnes$^3$, Johannes Aprojanz$^1$, Stephen R. Power$^3$, Antti-Pekka Jauho$^3$, Christoph Tegenkamp$^{1,2}$} \email{tegenkamp@fkp.uni-hannover.de}

\affiliation{$^1$ Institut f\"ur Festk\"orperphysik, Leibniz Universit\"at
Hannover, Appelstra\ss e 2, 30167 Hannover, Germany}
\affiliation{$^2$ Laboratory of Nano and Quantum Engineering (LNQE), Leibniz Universit\"at Hannover, Schneiderberg 39, 30167 Hannover, Germany}
\affiliation{$^3$ Center for Nanostructured Graphene, DTU Nanotech, Technical University of Denmark, 2800 Kgs. Lyngby, Denmark}


\begin{abstract}
We have realized nanometer size constrictions in ballistic graphene nanoribbons grown on sidewalls of SiC mesa structures. The high quality of our devices allows the observation of a number of electronic quantum interference phenomena.
The transmissions of  Fabry-Perot like resonances were probed by in-situ transport measurements at various temperatures. The energies of the resonances  are determined by the size of the constrictions  which can be controlled precisely using STM lithography. The temperature and size dependence of the measured conductances are in quantitative agreement with tight-binding calculations.
The fact that these interference effects are visible even at room temperature makes the reported devices attractive as building blocks for future carbon based electronics.
\end{abstract}
\maketitle

Graphene nanoribbons (GNRs) are an ideal system to study electronic transport phenomena in the coherent regime due to the extremely long mean free path and coherence length of charge carriers \cite{Baringhaus14, Berger06, Mayorov11}. In analogy with subwavelength optics, the coherent transmission of electrons through narrow constrictions within such ballistic ribbons gives rise to interference phenomena \cite{Darancet09, Ihnatsenka12}. Graphene nanoconstrictions (GNCs) are an important building block in carbon electronics, especially for valleytronic applications \cite{Rycerz07}, and hence their atomically precise synthesis as well as electronic characterization is of great importance \cite{Tombros11, Munoz-Rojas06, Ozyilmaz07, Lu11}. Unfortunately, lithographically defined GNCs and GNRs exhibit rather rough edges and the inherent defect potentials limit drastically the achievable mean free paths \cite{Han10, Todd09}.

The growth of graphene on the sidewalls of SiC mesa structures
was reported to produce graphene nanostructures of exceptionally high quality \cite{Sprinkle10, deHeer11, Nevius14, Hicks13, Baringhaus14, Baringhaus13a, Baringhaus15, Palacio15}. Their hallmark feature is the ballistic transport of electrons which can be observed on $\mu$m length scales \cite{Baringhaus14, Baringhaus15}. The robustness of the ballistic behavior makes these devices a prime platform for studying interference phenomena at graphene interfaces. For the patterning of narrow constrictions into the sidewall ribbon, STM lithography is the method of choice. It was shown to cut graphene sheets with atomic precision while preserving
the quality of the pristine material away from the cut \cite{Magda14, Tapaszto08}.

\begin{figure}
\includegraphics[width=1\columnwidth]{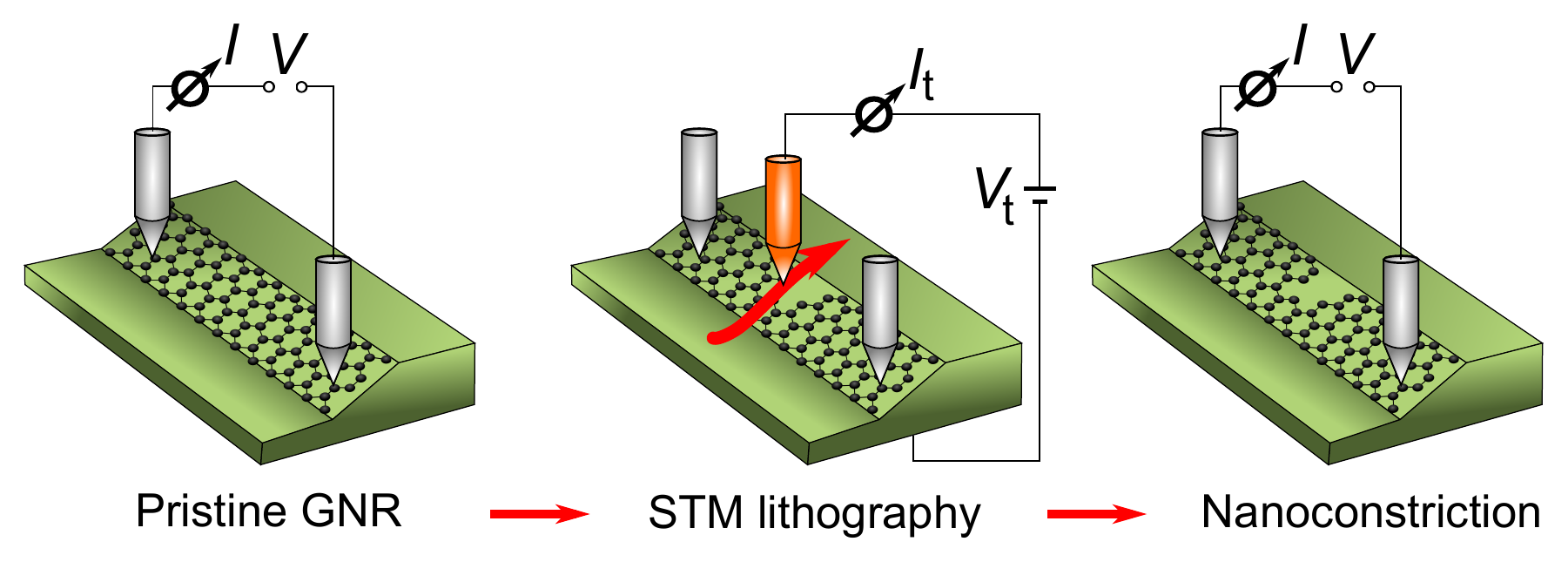}
\caption{\label{fig1} Synthesis of sidewall GNCs by STM lithography. First, a reference $2$-point-probe measurement is performed on the pristine sidewall GNR to ensure the presence of a ballistic channel. The GNC is subsequently defined into the GNR via local etching by means of a STM tip. A second $2$-point measurement probes the transport properties of the GNC.}
\end{figure}

For the growth of GNRs we use SiC wafers commercially purchased from SiCrystal AG. SiC substrates were flattened by using the face-to-face heating method \cite{Yu11, Baringhaus15} and subsequently mesa structures with lateral dimensions between $1\,\mu\mathrm{m}$ and $8\,\mu\mathrm{m}$ and a height of $20\,\mathrm{nm}$ were defined by using standard UV lithography combined with reactive ion etching (gas mixture 20/7 SF$_6$/O$_2$, power $30\,\mathrm{W}$). GNRs were grown exclusively on the sidewall of the mesa following standard recipes \cite{Sprinkle10, Baringhaus15}.

\begin{figure}
\includegraphics[width=\columnwidth]{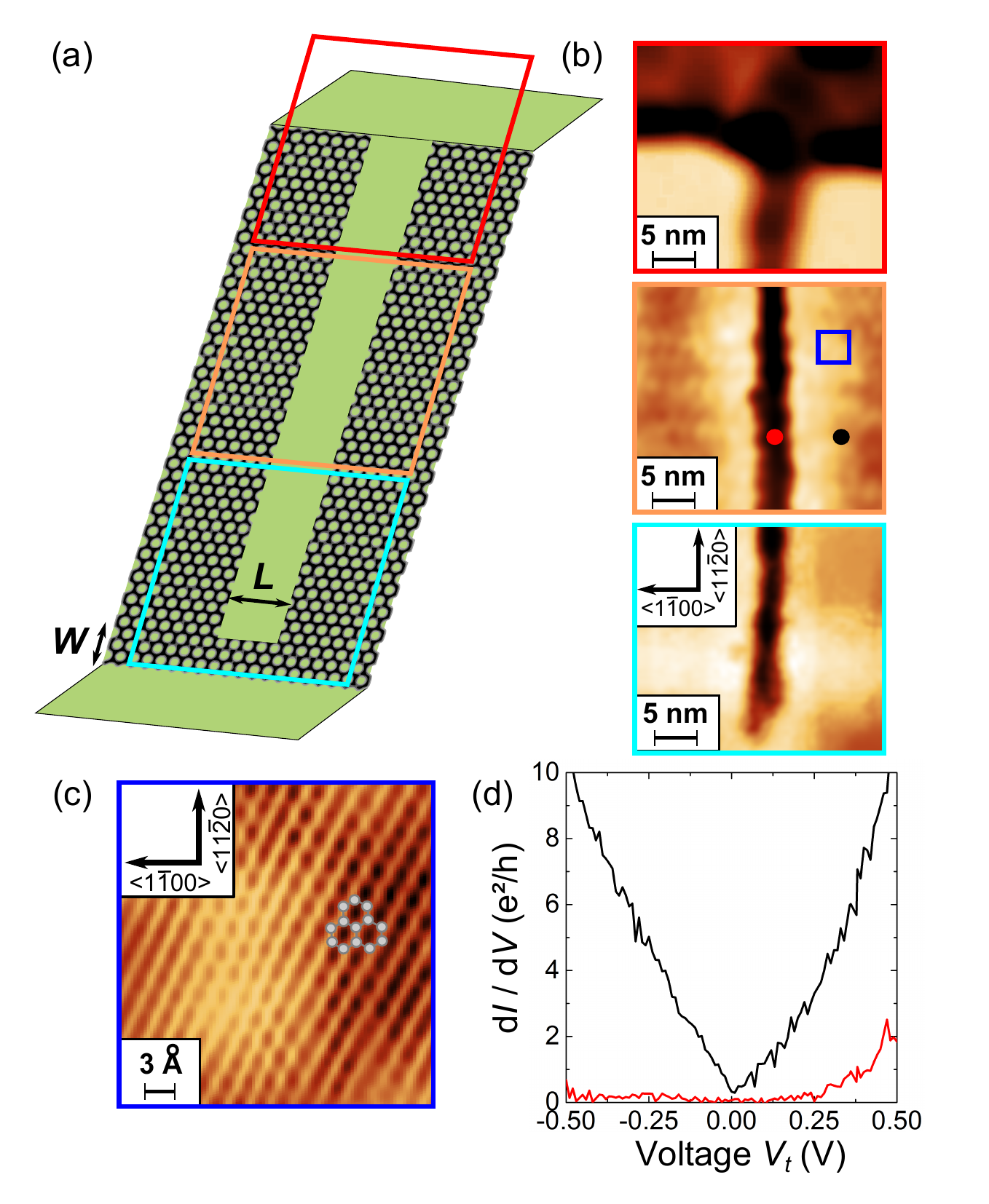}
\caption{\label{fig2}STM study of a sidewall GNC. a) Schematic view of the graphene structure on the SiC mesa sidewall after synthesis of a narrow constriction with STM lithography. b) STM images of the top, center and bottom part of the cut (tip voltage for imaging $V_\mathrm{t}=3\,\mathrm{V}$, tunneling current $I_\mathrm{t}=1\,\mathrm{nA}$). In the bottom, the presence of a GNC is confirmed.  The red and black dots in the middle frame indicate the locations at which spectroscopy was performed, panel (d). c) Atomically resolved STM topograph of the graphene lattice in the vicinity of the cut ($V_\mathrm{t}=200\,\mathrm{mV}$, $I_\mathrm{t}=100\,\mathrm{pA}$). d) $\mathrm{d}I_\mathrm{t}/\mathrm{d}V_\mathrm{t}$ spectroscopy of the GNC. Red curve: Spectrum acquired directly on the cut (open-feedback parameters (setpoint): $V_\mathrm{t}=0.2\,\mathrm{V}$, $I_\mathrm{t}=0.2\,\mathrm{nA}$, modulation voltage $V_\mathrm{rms}=15\,\mathrm{mV}$). Black curve: Spectrum acquired in the vicinity of the cut on the unpatterned graphene lattice ($V_\mathrm{t}=0.4\,\mathrm{V}$, $I_\mathrm{t}=0.4\,\mathrm{nA}$, modulation voltage $V_\mathrm{rms}=20\,\mathrm{mV}$).}
\end{figure}

A 4-tip STM in combination with a high-resolution SEM is used for both transport experiments and in-situ patterning of the nanoribbons. The local characterization of graphene and ballistic sidewall graphene nanoribbons with multiple STM probes is non-destructive and highly controlled \cite{Baringhaus14, Baringhaus13, Baringhaus15} and offers a unique possibility to study directional transport effects in graphene nanostructures \cite{Settnes14}. Further details about the experiments as well as the theoretical modeling are explained in the Supplemental Material \cite{RefSM}.

The most intriguing feature of the ballistic sidewall ribbons is a probe-spacing and temperature independent conductance of $1\,e^2/h$ \cite{Baringhaus14} which indicates single-channel transport. In such a ballistic ribbon, abrupt graphene interfaces can be introduced in the form of a narrow, a few nm wide and long, constriction. For this purpose, a STM tip is navigated across the graphene covered sidewall under extreme tunneling conditions, i.e. at large bias voltages and tunneling currents of about $V_\mathrm{t}<-5\,\mathrm{V}$ and $I_\mathrm{t}\geq50\,\mathrm{nA}$. The graphene underneath the STM tip is thereby removed (cf. \cite{RefSM}, Fig. 1).  The underlying etching mechanism is not fully understood, but relies most likely on the local breaking of carbon-carbon bonds underneath the STM tip via field-emitted electrons \cite{Venema97, Kim03}. The tip was always moved transversely over the ribbon starting at the trench and ending on the plateau of the mesa structure as shown schematically in Fig.~\ref{fig1}. This results in the formation of a constriction at the lower edge of the ribbon as  verified by  a subsequent STM characterization. For etching away the graphene at the lower ribbon edge, much higher etching voltages are needed. This is most likely due to the different geometry of this edge which terminates almost vertically into the substrate \cite{Baringhaus14, Norimatsu10}.

A schematic view of the graphene nanoconstriction obtained by STM lithography is given in Fig.~\ref{fig2}a. The corresponding STM images of characteristic positions along the cut are displayed in Fig. \ref{fig2}b showing the lower ribbon edge, the central part and the upper edge. The graphene appears bright compared to the underlying substrate. Obviously, the sidewall GNR was cut through at the upper edge to the mesa plateau and the middle of the ribbon, but not at the lower edge to the trench. Here, a small patch of graphene remains, forming a narrow constriction with lateral dimensions of only a few nm. Atomically resolved STM images ensure that the graphene lattice in the vicinity of the cut was not damaged by the cutting procedure as shown in Fig.~\ref{fig2}c.   The local density of states (LDOS), obtained by scanning tunneling spectroscopy (STS), further supports this finding. Tunneling spectra, Fig.~\ref{fig2}d,  were taken directly on the cut and on the surrounding graphene (Fig.~\ref{fig2}c, middle frame, red and black dot, respectively). The LDOS of the intact graphene surrounding the cut exhibits the characteristic V shape, with a tunneling conductance minimum precisely at $0\,\mathrm{V}$. Hence, the charge neutrality of the sidewall ribbon  is preserved \cite{Baringhaus14, Baringhaus15}. On the other hand, the LDOS recorded on the cut, drops to zero for the tunneling bias voltage range $-0.50\,\mathrm{V}<V_\mathrm{t}<0.25\,\mathrm{V}$. Hence, an electronic current flowing from the left to the right has to be transmitted through the constriction.

\begin{figure*}
\includegraphics[width=.9\textwidth]{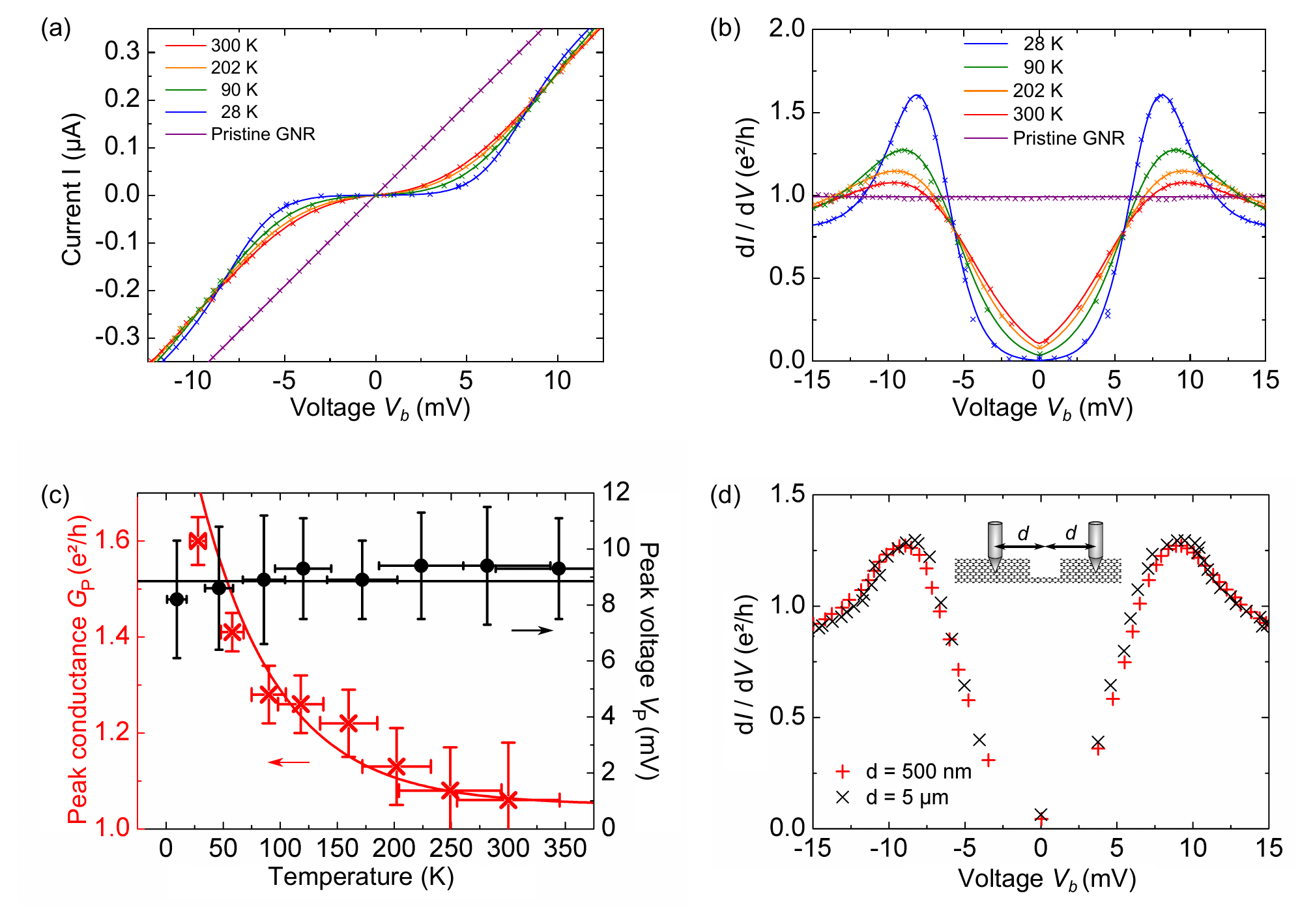}
\caption{\label{fig3} Electronic transport across GNCs. a) IV curves across a GNC with length $L=6\,\mathrm{nm}$ and width $W=2\,\mathrm{nm}$ for temperatures between $T=28\,\mathrm{K}$ and $300\,\mathrm{K}$. The solid lines indicate corresponding fits to the Kaiser expression \cite{Kaiser05}. The violet curve is the IV of the pristine ribbon, prior to STM lithography. b) Differential conductance of the IV shown in  a). Conductance peaks $G_\mathrm{P}$ at voltage $V_\mathrm{P}$ are clearly visible for the complete temperature range. The violet curve is the differential conductance of the pristine ribbon. c) Peak conductances $G_\mathrm{P}$ and peak voltages $V_\mathrm{P}$ extracted from differential conductance curves. The peak conductance is exponentially decreasing with increasing temperature while the voltage position of the peak remains constant. d) Differential conductance across a GNC measured with two different contact spacings $d=500\,\mathrm{nm}$ and $5\,\mu\mathrm{m}$. The inset depicts the arrangement of the probes on both sides of the constriction.}
\end{figure*}

The electronic transport through the sidewall GNC was recorded by biasing the constriction with two tips placed in ohmic contact on the sidewall GNR with the GNC in between (as schematically indicated in Fig.~\ref{fig1}). The IV characteristics of a biased sidewall GNC ($L=6\,\mathrm{nm}$, $W=2\,\mathrm{nm}$) recorded at different temperatures in the range from $28\,\mathrm{K}$ to $300\,\mathrm{K}$, are shown in Fig.~\ref{fig3}a  together with a reference measurement of the pristine, unpatterned sidewall GNR. The pristine ribbon exhibits a completely linear IV with a conductance of $1\,e^2/h$ which is characteristic for ballistic transport in a fully non-degenerate channel \cite{Baringhaus14}. In contrast, the IV through the constrictions are clearly nonlinear. They can be well described by the  phenomenological Kaiser expression \cite{Kaiser05} (details are given in \cite{RefSM}, Sect. 2) which is frequently used to describe nonlinearities in the IV-curves for carbon nanotubes. In the low-bias regime, the opening of a small transport gap ($\Delta\approx10\,\mathrm{meV}$) is clearly visible. For $T=28\,\mathrm{K}$ the zero-bias conductance drops to zero. With increasing temperature, the slope around zero bias increases while for $V_\mathrm{b}>10\,\mathrm{meV}$ it remains almost constant throughout the whole temperature range.

More insight can be gained from the differential conductance displayed in Fig.~\ref{fig3}b. The $\mathrm{d}I/\mathrm{d}V$  curves were obtained by either the numerical differentiation of the IV curves and averaging over at least $50$ individual measurements, or directly by using standard 
low-frequency lock-in techniques. The results of both methods agree well with each other (cf. \cite{RefSM}, Fig. 3) and are not distinguished in the following. 
In order to avoid Joule heating in these constrictions most of the experiments were performed with current densities not exceeding $10^8 \mathrm{A/cm^2}$.
The differential conductance curves of the sidewall GNC clearly indicate the opening of a transport gap around zero bias accompanied by conductance peaks located symmetrically at about $\pm9\,\mathrm{mV}$. The  positions of the conductance peaks ($V_\mathrm{P}$) remain almost constant throughout the whole temperature range and show only a slight decrease in the low temperature regime. The conductance peaks are most prominent at low temperatures and decrease with increasing $T$ in an exponential manner $G_\mathrm{peak}\sim\mathrm{exp}(-k_\mathrm{B}T/eV_0)$ as shown in Fig.~\ref{fig3}c. However, the peaks remain visible even up to room temperature. The maximum peak value is reached at the lowest attainable temperature $T=28\,\mathrm{K}$ with $G_\mathrm{P}\approx1.6\,e^2/h$. The peak conductance is significantly higher than the conductance of the pristine ribbon which indicates the opening of a second transport channel or the restoring of spin degeneracy in the whole ribbon. This result is surprising since dual channel transport was found in sidewall GNRs only for contact spacings well below $500\,\mathrm{nm}$ \cite{Baringhaus14, RefSM, RefSM7, RefSM8, RefSM9, RefSM10, RefSM11}. Here, we used contact spacings $d \geq 500\,\mathrm{nm}$ and hence expect no contribution from the second transport channel. The characteristic shape of the $\mathrm{d}I/\mathrm{d}V$ originates solely from the constriction itself and is independent of the contact spacing. This can be directly seen from Fig.~\ref{fig3}d where two conductance curves for two different contact spacings ($d=500\,\mathrm{nm}$ and $d=5\,\mu\mathrm{m}$) are displayed. The contact spacing has no influence neither on the occurence or amplitude of the conductance peak nor the opening of a transport gap.

\begin{figure}
\includegraphics[width=\columnwidth]{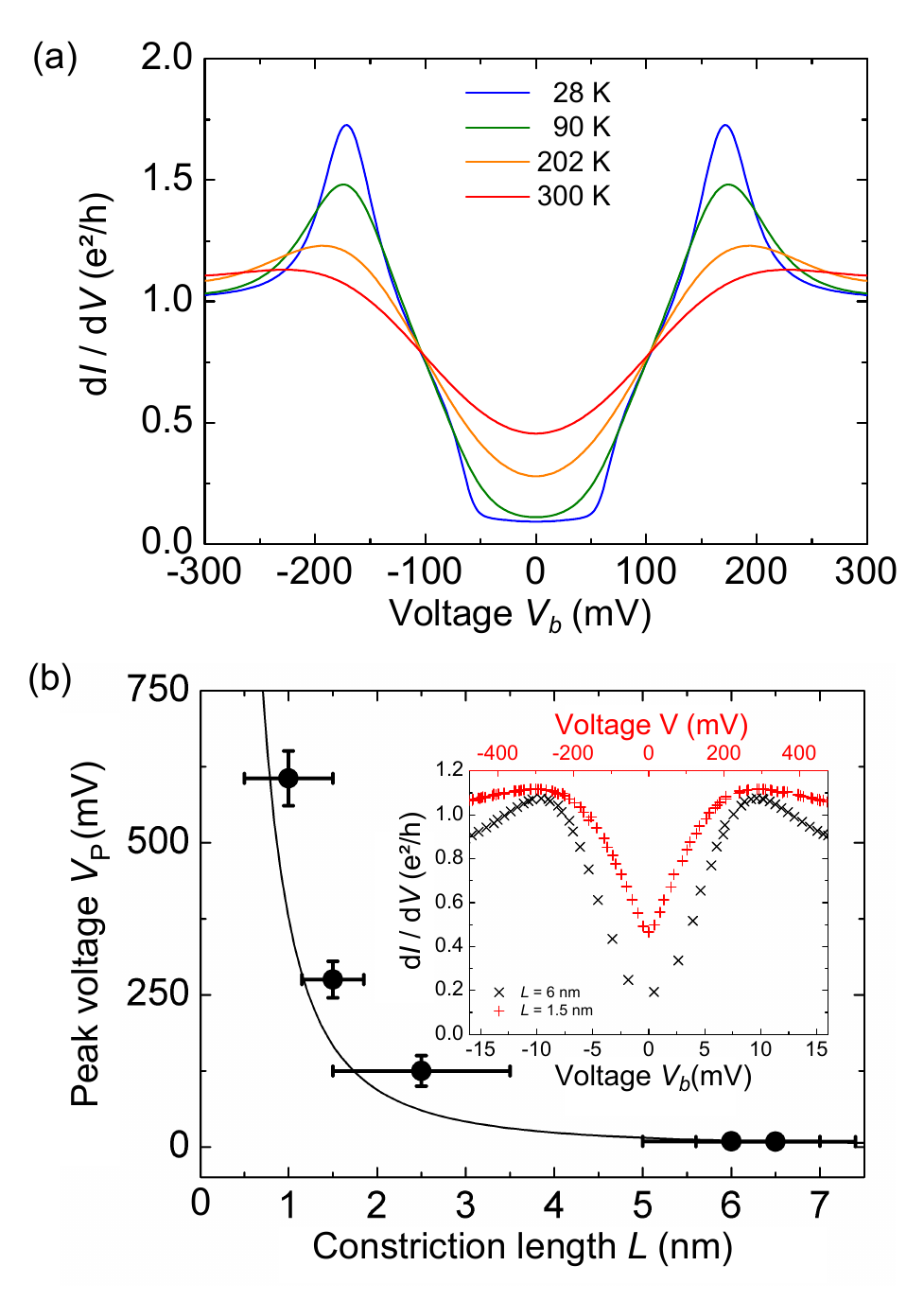}
\caption{\label{fig4} Theoretical transport characteristics of GNCs and comparison with the experiment. a) Theoretical differential conductance of a GNC of dimension  $L\approx1\,\mathrm{nm}$ and $W\approx1.7\,\mathrm{nm}$ for different temperatures.
b) Voltage position of the conductance peaks plotted against the length $L$ of the constriction. The solid black line indicates the energy of the ground state in the standard particle-in-a-box picture. Inset: Measured conductance at $T=298$ K for constriction lengths $L\approx 6$ nm (black) and $L\approx 1.5$ nm (red).  Note the different voltage scales for the two curves.}
\end{figure}

In order to explain the opening of a transport gap as well as the origin of the conductance peaks, the GNC can be viewed as a diffraction barrier. Electronic diffraction gives rise to localized currents through the constriction which subsequently lead to transmission resonances \cite{Ihnatsenka12, Gunst13}. In analogy to subwavelength optics, the whole system can be treated as a Fabry-Perot cavity \cite{Darancet09}. For a more quantitative analysis, we calculate the bias dependence of the differential conductance through a GNC with a propagating edge state. In this way, we model the GNC system by a third nearest neighbor tight binding model using a recursive Green's function approach within the common Landauer-B\"{u}ttiker formulation (see \cite{RefSM, RefSM1,RefSM2, RefSM3, RefSM4, RefSM5, RefSM6} Sect. 4, where also effects due to disorder are discussed.). Fig.~\ref{fig4}a shows a typical result for a constriction with dimensions $L \approx 1\,\mathrm{nm}$ and $W \approx 1.7\,\mathrm{nm}$ at various temperatures. Pronounced conduction peaks are  found in the simulations, resembling nicely the experimental findings.   Thereby the interpretation that the constrictions give rise to resonance phenomena is strongly supported. With increasing temperature we notice that the peaks broaden and the peak heights decrease in a similar manner as observed experimentally.
We also want to emphasize that for our GNC configuration the resonances originate from the zeroth mode of the zigzag-topology of the ribbons, thus mode coupling effects, e.g. coming along with anti-resonances like seen in wider 2DEG wire systems, is not pivotal in our case within the small energy window \cite{Nakazato91}.

The edge state gives rise to the constant $1\,e^2/h$ conductance regime for $V > V_\mathrm{P}$, whereas the peak features are caused by resonances within the constriction. The energy of the resonance is highly dependent on the length of the constriction which can be understood in terms of a simple particle-in-a-box picture. Indeed, the resonance energy follows accurately the scaling law $E \propto 1/L^2$ as shown in Fig.~\ref{fig4}b. While the voltage position of the resonance peak shifts drastically upon changing the dimensions of the constriction, the general shape of the differential conductance remains almost unaltered (cf. inset of Fig.~\ref{fig4}b where the $\mathrm{d}I/\mathrm{d}V$ curves at $T=298\,\mathrm{K}$ are shown for constrictions of length $L \approx 6\,\mathrm{nm}$ (black) and $L \approx 1.5\,\mathrm{nm}$ (red)). This indicates that the underlying physical mechanism which gives rise to the resonance peaks is the same for narrow and wide constrictions.

These results show that graphene nanoconstrictions can be effectively used as electronic diffraction barriers. GNRs on the sidewalls of SiC mesa structures are an ideal template for this purpose. STM lithography allows us to define constrictions in-situ with variable dimensions of only a few nm. The excellent structural and electronic quality of the self-assembled ribbons as well as the subsequently defined constrictions gives rise to electron interference phenomena. Direct probing with local transport reveals the emergence of conductance peaks and transport gaps which can serve as hallmark for electron interference. The stability of these features up to room-temperature opens up the possibility to use them in novel electronic nanodevices. Sidewall GNRs can serve as connectors between multiple constrictions as well as to the contacts over distances of several $\mu$m due to their exceptional transport properties. Hence, such devices would solely rely on the photon-like nature of electrons in graphene and belong to a new class of fully coherent electronics.

\vspace{1cm}
{\bf Acknowledgement}\\
Financial support by the Deutsche Forschungsgemeinschaft is gratefully acknowledged by J.B., J.A. and C.T..
The Center for Nanostructured Graphene (CNG) is sponsored by the Danish National Research Foundation, Project No. DNRF103.

%
%
%
%
%
%
%
%
%
%
%
%

\pagebreak
\widetext
\begin{center}
	\textbf{\large Supplemental Materials: Electron interference in ballistic graphene nanoconstrictions}
\end{center}
\setcounter{equation}{0}
\setcounter{figure}{0}
\setcounter{table}{0}
\setcounter{page}{1}
\makeatletter
\renewcommand{\theequation}{S\arabic{equation}}
\renewcommand{\thefigure}{S\arabic{figure}}
\renewcommand{\bibnumfmt}[1]{[S#1]}
\renewcommand{\citenumfont}[1]{S#1}

	\noindent The supplementary material contains the following:
\begin{itemize}
	\item[1.] Details about STM measurements ans local etching of graphene with STM lithography
	
	\item[2.] Details about the fitting of IV curves using the Kaiser expression
	
	\item[3.] Comparison of differential conductances obtained numerically and by Lock-in techniques
	
	\item[4.] Calculation of the differential conductance through GNCs via tight-binding
	
	\item[5.] Ballistic transport channels in sidewall GNRs grown on 6H-SiC(0001)
	
\end{itemize} 
	
	\section{STM and STM lithography}
	
	We use an Omicron Nanoprobe system for all transport and STM experiments. It is equipped
	with four individual STM tips and a high-resolution Gemini SEM for tip placement. All STM
	images were recorded at room temperature. Tunneling spectra were recorded using standard lock-in
	techniques in open-feedback configuration.
	Contacts to the GNRs for transport experiments were realized with two of the STM tips
	placed in direct, ohmic contact onto the graphene. Tips were first brought into tunneling contact
	and driven to their desired positions. After switching of the feedback, the tips were lowered to the
	sample surface while checking the contact resistance until stable contact is reached. All transport
	experiments were done in a two-point configuration with electrochemically etched tungsten tips.
	
	The local etching of graphene in UHV was accomplished by applying high bias voltages ($V_\mathrm{t}<-5\,\mathrm{V}$) to the tungsten STM tip while simultaneously moving it slowly ($\approx 2 \mathrm{nm}/\mathrm{s}$) over the sample. After the cutting procedure, the structure was imaged again with the same STM tip. As an example, Fig. S1 shows an STM topograph of three cuts obtained with cutting voltages of $-6\,\mathrm{V}$, $-7\,\mathrm{V}$ and $-8\,\mathrm{V}$. In general, the best etching results were obtained with negative tip voltages.
	
	\begin{figure}[b!]
		\includegraphics[width=0.3\textwidth]{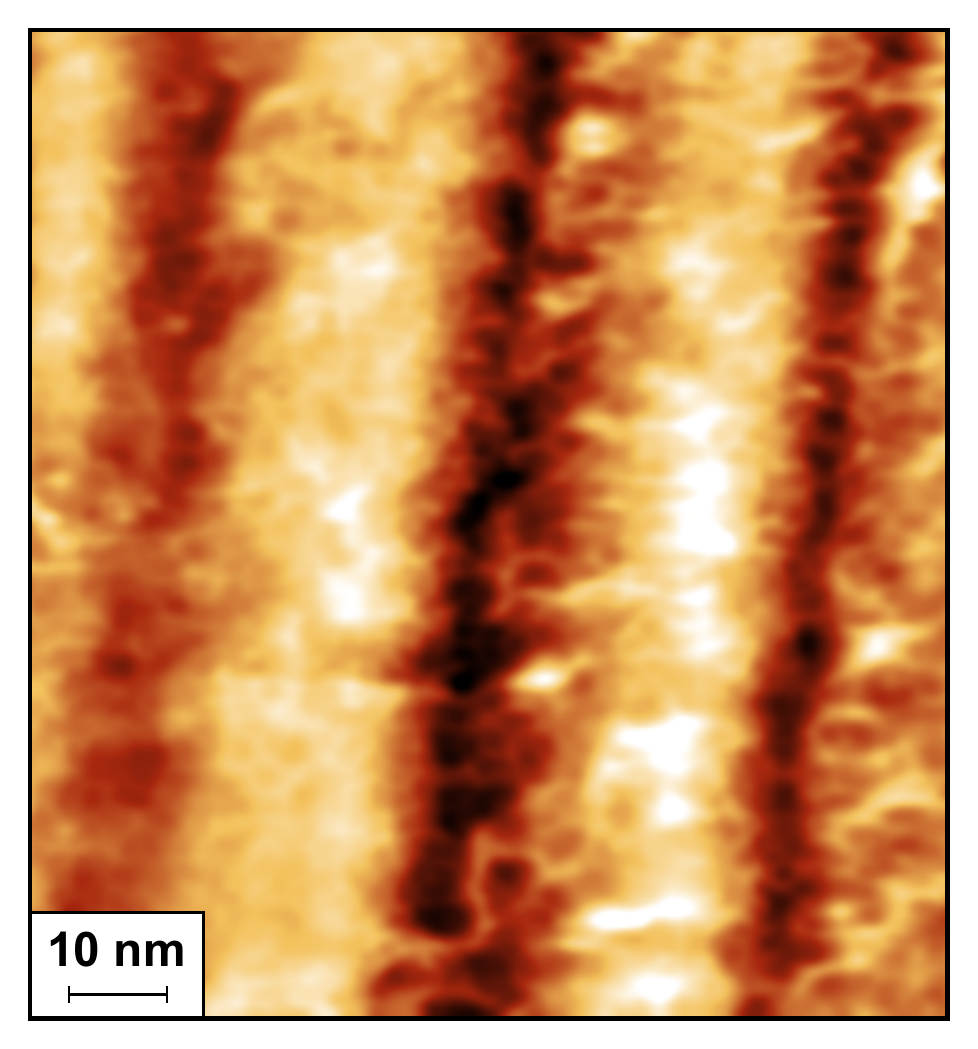}
		\caption{\bfseries STM lithography on planar graphene sheets  \mdseries  STM topography image ($V_\mathrm{t}=500\,\mathrm{mV}$, $I_\mathrm{t}=500\,\mathrm{pA}$) of three etching lines written with STM bias voltages of (from left to right) $-6\,V$, $-7\,V$ and $-8\,V$.}
		\label{figS5}
	\end{figure}
	
	\section{Kaiser expression}

	The Kaiser expression \cite{Kaiser05_Supp} can be used to describe the IV characteristics of the GNCs phenomenologically. It is a generic expression frequently used to describe nonlinear voltage characteristics of carbon nanotubes or nanofibres. It can be expressed as \cite{Kaiser05_Supp}
	
	\begin{equation}
	\frac{I}{V} = \frac{G_0 \mathrm{exp}(\frac{V}{V_0})}{1+ \frac{G_0}{G_\mathrm{h}} (\mathrm{exp}(\frac{V}{V_0})-1)}
	\label{Kaiser_expression}
	\end{equation}
	
	\noindent where $G_0$ denotes the conductance for $V \rightarrow 0\,\mathrm{V}$, $G_\mathrm{h}$ is the saturation conductance at large bias and $V_0$ is a voltage scale factor. The good agreement of this expression with the IV data from GNCs is obvious from Fig. S2a. The description with this generic function allows to easily determine the differential conductance by simple derivation of the fit function. Fig. S2b shows the resulting differential conductance. Additionally the differential conductance values extracted from the raw data IV curves by numerical derivation are shown as data points.
	
	\begin{figure}
		\includegraphics[width=1\textwidth]{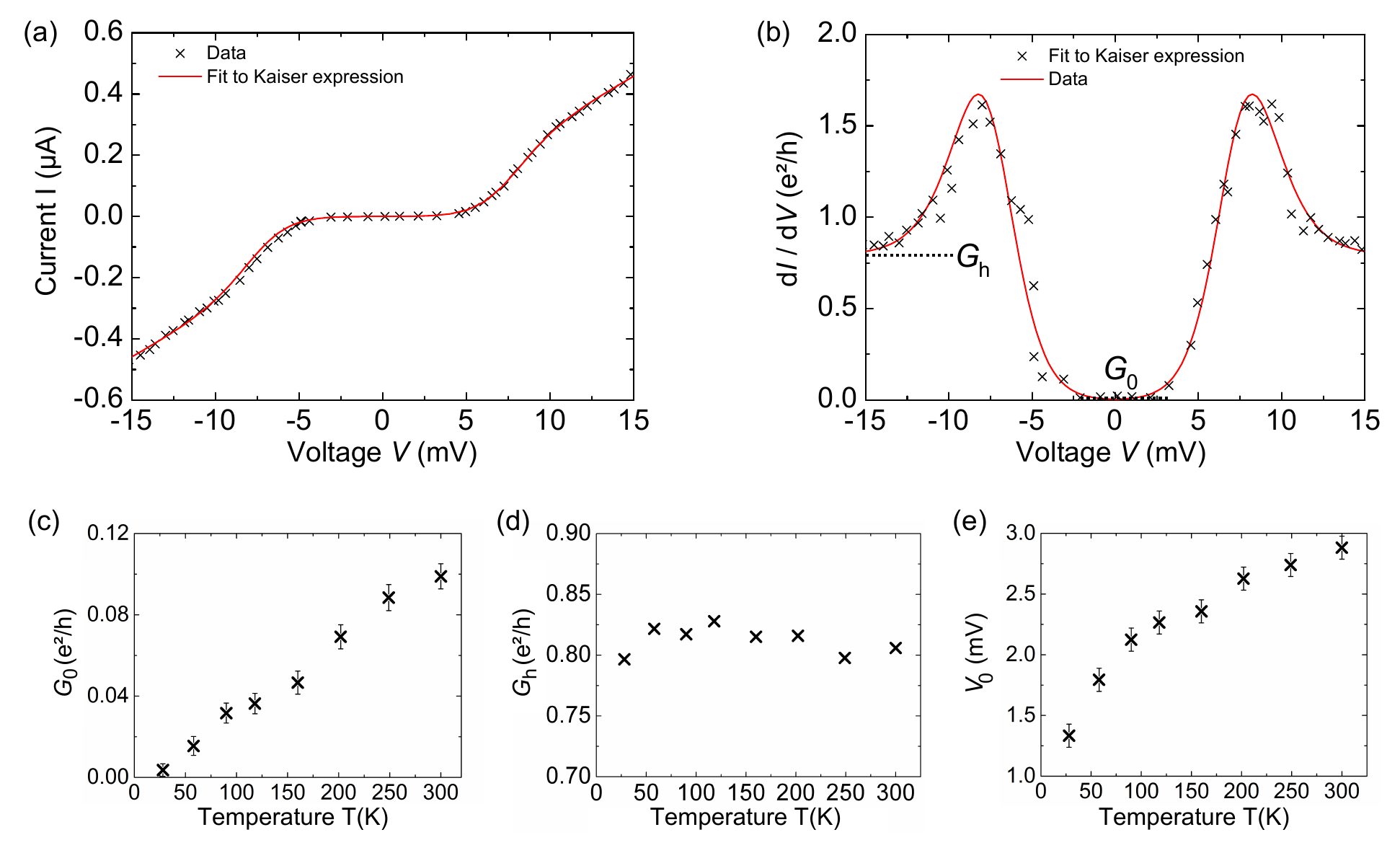}
		\caption{\bfseries Fitting of transport data with the Kaiser expression.  \mdseries \bfseries a \mdseries IV curve across a GNC with length $L=6\,\mathrm{nm}$ and with $W=2\,\mathrm{nm}$ at $T=28\,\mathrm{K}$. The red line indicate a corresponding fit to the Kaiser expression. \bfseries b \mdseries Differential conductance of both the experimentally obtained IV and the fit shown in \bfseries a \mdseries.}
		\label{figS1}
	\end{figure}
	
	The parameters, used for fitting the individual IV curves to the Kaiser expression, are plotted against temperature in Fig. S2c-e. The zero bias conductance $G_0$ increases with increasing temperature in the same manner as the voltage scale factor $V_0$. The high bias conductance $G_\mathrm{h}$ of about $0.8\,e^2/h$ is almost constant throughout the whole temperature range.

	\section{Lock-In vs. Numerical derivation}
	
	The differential conductance curves presented in the paper as well as the supplemenent were either obtained by the numerical derivation of measured IV curves or directly by using low-frequency Lock-In techniques. In order to reduce the noise level in case of the numerical derivation, the average of at least 50 individual IV curves was used. Both methods, the Lock-In technique and the numerical derivation, lead to comparable results as exemplarily shown in Fig. S3. The shape of the curve is almost identical and especially, the conductance peaks around $\pm9\,\mathrm{mV}$ are clearly visible and seen at the same position with both methods.
	
	\begin{figure}
		\includegraphics[width=0.5\textwidth]{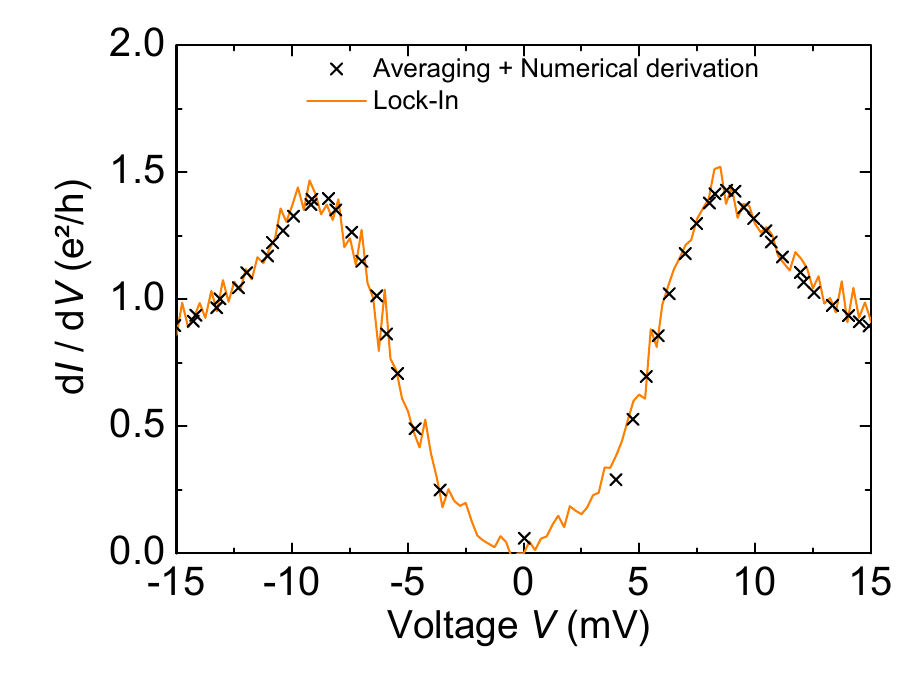}
		\caption{\bfseries Comparison of differential conductances obtained by numerical derivation and Lock-In techniques. \mdseries Data obtained by numerical derivation were gained after averaging 50 individual IV curves. Lock-In data were recorded using standard low frequency Lock-In techniques. Both data sets were recorded at $T=40\,\mathrm{K}$.}
		\label{figS2}
	\end{figure}
	
	\section{Tight-binding calculations}
	
	To analyze the differential conductance, we consider a simple model of a constriction in a nanoribbon containing a propagating edge state. \cite{Chang2012_Supp}
	We employ a third nearest neighbor tight binding scheme using the Hamiltonian
	\begin{align}
	\mathbf{H} = - \sum_{k=1}^{3} \sum_{<i,j>} \gamma_{ij}^k c_i^\dagger c_j + h.c,
	\end{align}
	where the sums run over first, second and third nearest neighbor pairs using the hopping parameters $\gamma^1 = 3.1$ eV, $\gamma^2 = 0.2$ eV and $\gamma^3 = 0.16$ eV. \cite{Lherbier2012_Supp}
	We consider a graphene nanoconstriction as shown schematically in Fig. S4a.
	The width of the ribbon hosting the constriction is chosen to be $N_y \approx 13$ nm.
	We note that the width of the hosting ribbon determines only the sharpness of the resonance peaks stemming from the constrictions and not their positions in energy.
	Furthermore, as the width of the hosting ribbon is increased, the energy onset of higher order modes is decreased and for very wide ribbons, these higher order modes are superimposed on the resonance phenomena.
	As the higher order modes are washed out in the experimental setup due to the long probe separation, it is sufficient to focus on a GNC system where the onset of the higher order modes is outside the energy range considered for the resonance effects. Further, we focus solely on the first resonance peak observed in the transmission spectrum, as higher order resonance peaks follows the standard particle in a box argument by being at more than twice the energy of the first resonance and therefore falls outside the considered energy interval \cite{Darancet09_Supp}.
	
	\begin{figure}
		\includegraphics[width=1\textwidth]{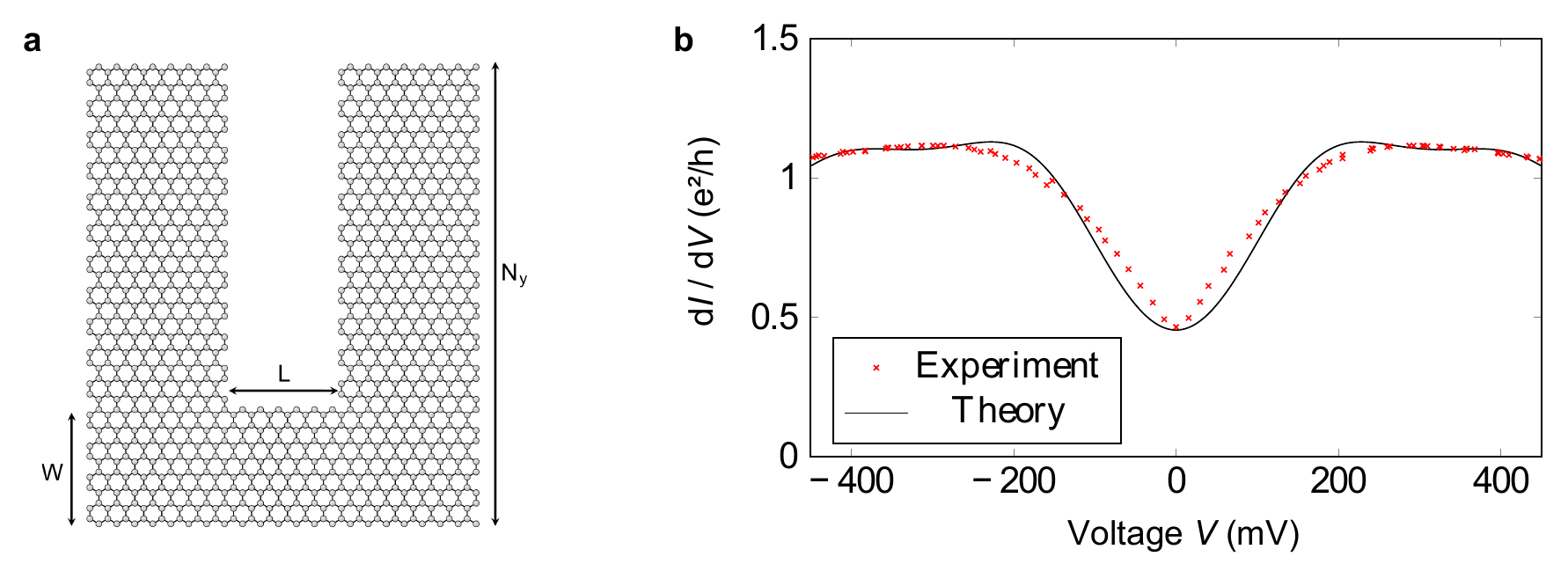}
		\caption{\bfseries Comparison with experimental data.  \mdseries \bfseries a \mdseries Sketch showing a graphene nanoconstriction of width $W$ and length $L$. The shown constriction is not the actual size used in the calculations. The constriction is placed in a pristine nanoribbon with zigzag edges of width $N_y$. \bfseries b \mdseries Comparing the experimental data for a constriction of width $W\sim 1.5$ nm and temperature $T=298$ K with the numerical calculation of $\mathrm{d}I/\mathrm{d}V$ at $T = 298$ K for a constriction with length $L\approx 1$ nm and width $W\approx 1.7$ nm embedded within a pristine nanoribbon of width $N_y \approx 13$ nm.}
		\label{fig:theo_supp_fig1}
	\end{figure}
	
	We calculate the current from \cite{DattaBook}
	\begin{align}
	I(V,T) = \frac{2\mathrm{e}^2}{h}\int\mathrm{d}E  \big[f\big(E-eV/2,T\big) - f\big(E+eV/2,T\big)\big]\mathcal{T}(E).
	\end{align}
	where $V$ is the applied bias voltage and $f(E,T)$ is the Fermi Dirac distribution at energy $E$ and temperature $T$
	\begin{align}
	f(E,T) = \bigg[\mathrm{e}^{E/k_B T} +1\bigg]^{-1}.
	\end{align}
	The transmission function $\mathcal{T}(E)$ is given by
	\begin{align}
	\mathcal{T}(E) =\mathrm{Tr}\big[\mathbf{\Gamma}_L(E) \mathbf{G}(E) \mathbf{\Gamma}_R(E) \mathbf{G}^\dagger(E) )\big]
	\end{align}
	where the broadenings are $\mathbf{\Gamma}_{L/R}(E) = \mathrm{i} \big(\mathbf{\Sigma}_{L/R}-\mathbf{\Sigma}^\dagger_{L/R}\big)$ with lead self-energies $\mathbf{\Sigma}_{L/R}$ calculated numerically using the decimation method of Ref. \cite{Sancho1985_Supp}.
	The Green's function is given by \mbox{$\mathbf{G}(E)=\big[E-\mathbf{H}-\mathbf{\Sigma}_L(E) -\mathbf{\Sigma}_R(E) \big]^{-1}$} and can be calculated from the widely used recursive Green's function techniques as explained in Refs. \cite{Lewenkopf2013_Supp} and \cite{Settnes2015_Supp}.

	Simulations of a GNC with dimensions $L\approx 1$ nm and $W\approx 1.7$ nm embedded in a hosting ribbon with $N_y \approx 13$ nm are shown for various temperatures in Fig, 4a in the main manuscript, where a qualitative agreement with experimental Fig 3b was noted.
	In  Fig. S4b here, we further plot the differential conductance for this constriction together with the experimental $\mathrm{d}I/\mathrm{d}V$ curve for a GNC of length $L\sim 1.5$ nm (same experimental data as plotted by the red symbols on the inset of Fig. 4b in the main manuscript).
	$T = 298$ K for both measurement and numerical calculation.
	We observe a good qualitative agreement which confirms that the experimental data is consistent with the formation of nanometer sized constrictions acting like a diffraction barriers.

	\begin{figure}
		\includegraphics[width=1\textwidth]{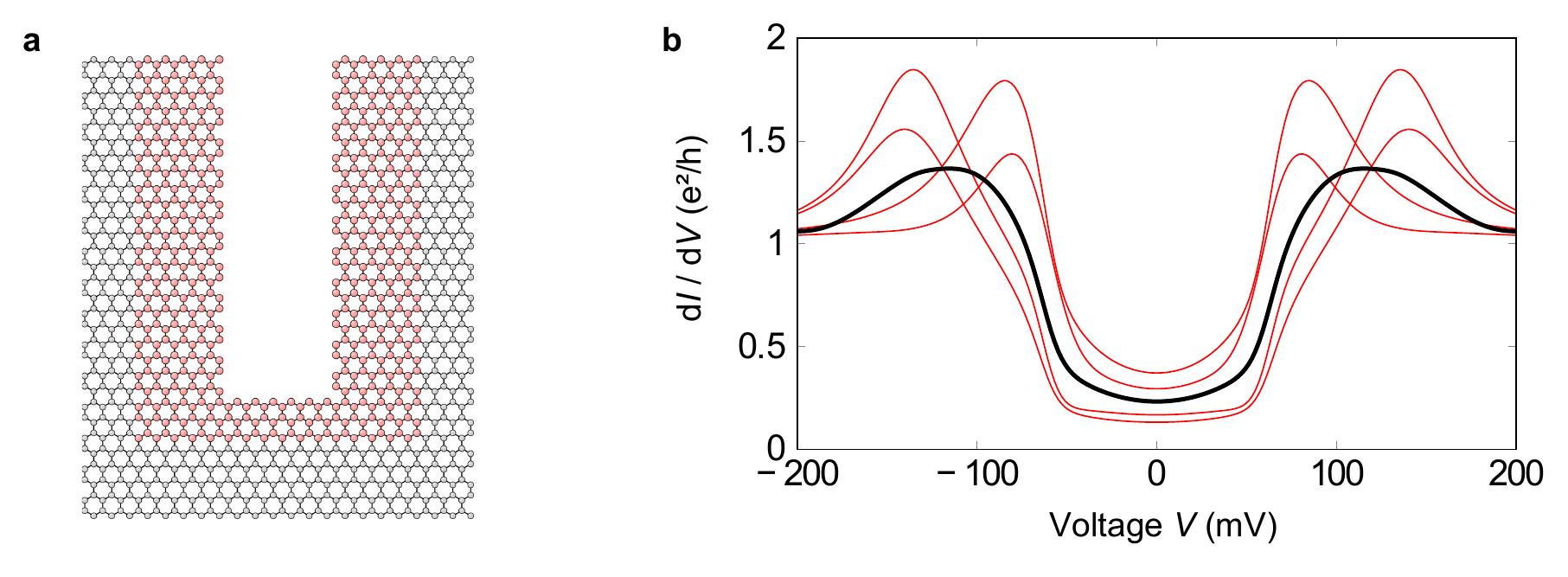}
		\caption{\bfseries Simulation of the effect of disorder on the transmission through a GNC.  \mdseries \bfseries a \mdseries Sketch showing the region with added disorder. \bfseries b \mdseries The differential conductance at $T = 28$ K for a constriction of length $L\approx 1.7$ nm and width $W\approx 1$ nm with Anderson disorder as described in the text. The black curve is the average of 25 individual disorder configurations and the red curves indicate a few representative realizations.}
		\label{fig:theo_supp_fig2}
	\end{figure}
	
	It is reasonable to assume that the STM lithography introduces disorder along the etched edge.
	Therefore, we study the resonance effects in the presence of random disorder along this edge.
	We consider Anderson type disorder, where the onsite energy is chosen randomly within the range $[-\delta W/2,\delta W/2]$ where $\delta W=|\gamma^1|/4$.
	We modify sites around the etched edge of the constriction as indicated by red sites in Fig. S5a.
	In Fig. S5b, we plot the average of several disorder realizations (black curve) together with a few individual disorder configurations (red curves).
	We notice that the peak feature is still clearly visible, although, we now observe a spread in the exact peak position.
	This suggests that the resonance features seen experimentally are robust against a considerable amount of disorder around the etched constriction edge.

	\section{Ballistic transport channels in sidewall GNRs grown on 6H-SiC(0001)}
	
	It was shown recently that sidewall graphene nanoribbons (GNRs) exhibit ballistic transport channels \cite{Baringhaus14_Supp}. In a follow-up study we were able to confirm that the existence of these transport channels come along with the formation of zigzag-type ribbons. Moreover, the roughness of the SiC(0001) surface is crucial and high terrace densities of the SiC host material easily change the mean free path lengths and, finally, even quench the ballistic transport channels. \cite{Baringhaus15_Supp}.
	
	Thereby,  the inherent 4-fold degeneracy of the states in is lifted in our GNR structures. A lift of the pseudo-spin degeneracy can be rationalized as the ribbons turn out be strongly buckled  in consequence of the growth of our ribbon on a SiC facet. Moreover, also the bonding of the edge easily can break the sublattice symmetry, thus  a residual conductance of $2e^2/h$ is expected and indeed seen for probe distances below 500~nm \cite{Baringhaus14_Supp}. A detailed probe-distance dependent analysis of the transport in the ballistic regime has revealed that the two transport channels are characterized by different mean free path lengths. For probe distances above 500~nm only one ballistic channel remains  and which we functionalized further in this study.
	
	Albeit  a conclusive model  for the crossover from a double-channel to a single-channel transport behavior is currently missing, there are first studies which point into the right direction. A recent paper by Chu and He \cite{Chu14_Supp} claims to explain our findings. It turns out that a $sp^3$ distortion of carbon atoms at the GNR edges induces a large spin-orbit coupling.  The formation of metallic edge states in zigzag ribbons has been confirmed by many calculations (see e.g. \cite{Huang13_Supp}). Thereby, the channels with same direction of propagation and opposite spin are located at opposite edges of the ribbon. In general, details of the interaction of the edges are crucial and easily  lift the spin-degeneracy.

	Indeed, transmission electron microscopy showed that particularly the lower edge of the ribbon is covalently bond to the lower SiC(0001) terrace \cite{Norimatsu10_Supp}. This can have severe implications to the electronic band structure. First of all, the vertical merging of the edge in to the SiC substrate is accompanied by in-place electric fields which were theoretically shown to affect the spin-texture of the states \cite{Son06_Supp}.  Moreover, curvature effects are present which are known to induce pseudo-magnetic fields \cite{Levy10_Supp}. Therefore, any kind of  effect, which breaks locally time reversal symmetry, enables backscattering (at the same edge) and may explain the probe distance behavior we have seen \cite{Baringhaus14_Supp}.
	
	Based on these findings and ideas, a detailed understanding for the crossover behavior will finally only succeed if the real environment of the GNR sidewall ribbons is taken into account.

\end{document}